\documentclass[epj]{svjour}
\usepackage{graphicx, epsfig}
\usepackage{times}
\usepackage[pdftex]{color}

\usepackage{amssymb, amsmath, bbm}
\usepackage[english]{babel} 
\usepackage{cite}

\newcommand{\unit}[1]{\,\mathrm{#1}}

\renewcommand{\vec}[1]{\mathbf{#1}}
\renewcommand{\Im}{\mbox{ Im }}
\renewcommand{\Re}{\mbox{ Re }}

\begin{document}
%
\title{Magnetic near fields as a probe of charge transport in spatially
dispersive conductors}
\author{H.\,R. Haakh \and C. Henkel
}                     
%
%
\institute{Institut f\"ur Physik und Astronomie, Universit\"at 
Potsdam, Karl-Liebknecht-Str. 24/25, 14476 Potsdam, Germany}
\date{Received: date / Revised version: date}
%
\abstract{
We calculate magnetic field fluctuations above a 
conductor with a nonlocal response (spatial dispersion) and consider
a large range of distances. The cross-over from ballistic to diffusive
charge transport leads to 
reduced noise spectrum at distances below the electronic
mean free path, as compared to a local description. 
We also find that the mean free path
provides a lower limit to the correlation (coherence) length of the
near field fluctuations.
The short-distance behavior is common to a wide range of materials,
covering also semiconductors and superconductors.
Our discussion is aimed at atom chip experiments where spin-flip transitions
give access to material properties with mesoscopic spatial resolution.
The results also hint at fundamental limits
to the coherent operation of miniaturized atom traps and matter wave 
interferometers.
} 
\maketitle
%




%






\section{Introduction}




%


 
Electromagnetic fluctuations have been playing a key role in physics 
ever since Planck
discovered the black-body spectrum. They have universal properties at distances
from a body large compared to the thermal (Wien) wavelength. 
In fact, the noise spectrum is telling a lot about material properties in the
near field, due to the links provided by the Kirchhoff law and
the fluctuation-dissi\-pation theorem 
\cite{Callen1951}. In practical applications like magnetic resonance
imaging, this near-field noise is a limiting factor for detecting biological
signals, for example \cite{Nenonen1996, Sidles2003}.
In the case of a metallic body, electromagnetic fluctuations
depend mainly on the conductivity and can reveal details about charge 
transport in the bulk. We focus here on a linear current-field relation 
in the general nonlocal form (spatial dispersion)
\begin{equation}
j_m(\vec{r}, \omega) = \sum_n\int d^3\vec{r'} \sigma_{mn}(\vec{r},\vec{r'}, \omega) 
E_n(\vec{r'},\omega)~,
	\label{eq:def-spatial-dispersion}
\end{equation}
where the dependence on both $\vec{r} $ and $\vec{r'}$ contains
the crossover from ballistic to diffusive in the motion of charge carriers.
This introduces the mean free path $\ell$ as a characteristic length scale.
The conductivity (or dielectric) tensor now depends on both frequency and 
wave-vector in Fourier space (\emph{spatial dispersion}) \cite{Reuter1948, 
Dingle1953a, Lindhard1954, Kliewer1968, Kliewer1969, Mermin1970, Ford1984}.
Further nonlocal effects are introduced by the details of the surface and 
the surface scattering of charge carriers \cite{Reuter1948, Ford1984, 
Kliewer1968, Kliewer1970, Foley1975, Flores1979,  Flores1977,Garcia1977, 
Feibelman1982}.

The anomalous skin effect is a famous consequence of the nonlocal bulk response. It is typically discussed in the regime of high frequencies, where the classical skin-depth $\delta(\omega)$ falls below the mean free path and does no longer describe screening of magnetic fields correctly \cite{Reuter1948, Pippard1947, Chambers1952}.
Other relevant physical phenomena are Thomas-Fermi (Debye-H\"uckel) screening and Landau damping \cite{Dressel2002}, connected to plasma screening due to mobile charges in the metal and electron-hole generation (internal photo effect), respectively. One can generally expect a reduction of field fluctuations close to the surface of a nonlocal metal as compared to local theory, because the bulk fields are better screened
and escape less easily into the surrounding space.
The nonlocal response of surfaces has also been discussed for other observables
outside the range of the anomalous skin effect. Its implications have been worked
out for the dispersion relation
of surface plasmon modes~\cite{Flores1979, Feibelman1982},
the transfer of heat via near field radiation~\cite{Volokitin2007, Chapuis2008}, 
the van der Waals interaction 
across a electrolyte \cite{Podgornik1987, Jancovici2006}, 
and the Casimir interaction between two metallic half-spaces \cite{Sernelius2005,Contreras-Reyes2005a,Esquivel2006}.

A motivation for this paper is the observation that 
spin-flip transitions of ultracold atoms 
held in miniaturized chip-based magnetic traps (atom chips) 
are sensitive to magnetic fluctuations in the
near field of a metal
\cite{Folman2002, Henkel2003, Fortagh2007, Reichel2011}. Here, atoms are probing surface 
properties at somewhat exotic frequencies in the radio or microwave band, much below the visible to ultraviolet frequency range of conventional metal 
spectroscopy. (Note, however, that the microwave band is routinely used
in superconductor experiments.) At the same time, 
the corresponding wavelengths are in the micron range because 
atoms illuminate the surface with their near field.
Modes characterized by these frequencies and wave vectors (parallel to the 
surface) lie in the evanescent sector, way below the light cone; in particular 
their in-plane wavevector $p$ is not restricted as in propagating vacuum fields. Therefore values $p \sim 1 / \ell$ 
cannot be excluded where spatial dispersion is clearly relevant.
Previous work \cite{Henkel1999a, Henkel1999, Jones2003, Rekdal2004} has 
considered spin-flip transitions for the microtrap scenario in the local limit, identifying them as relevant challenges to miniaturization below the
micron scale.
We also mention the 
results of Ref.\,\cite{Chklovskii1992} on nuclear spin relaxation 
that are reproduced and generalized here, including the skin effect and covering 
a wider range of distances.

Another motivation is the study of spatial correlations of thermal near field 
radiation. These differ strongly from the blackbody limit where the wavelength
provides a universal correlation (or coherence) 
length~\cite{MandelWolf}.
For an overview on the coherence
of thermal radiation see Ref.\,\cite{Greffet2007}. Electric field correlations 
were discussed previously in 
Refs.~\cite{Gori1994, Wolf2001, Blomstedt2007} for homogeneous media
and in Refs.~\cite{Carminati1999, Henkel2000, Dorofeyev2002, Henkel2006,
Lau2007, Norrman2011} for the near field of bodies. 
Surface charge and current
correlations have been studied in the high-temperature limit in
Refs.\,\cite{Boustani2006,Jancovici2006,Samaj2008}.
In the electric case, the mean free path $\ell$ did not emerge as
a characteristic length scale, neither in the distance dependence of the noise
spectrum nor in the spatial autocorrelation function~\cite{Henkel2006}.
This may be related to
sum rules and efficient screening at the surface. The magnetic field behaves
differently because at distances around the mean free path there is a crossover
in the noise spectrum~\cite{Chklovskii1992}. We show here that the
field correlations in a plane parallel to the metal surface become more coherent
at short distances ($z \ll \ell$), and that the correlation length involves the
mean free path.

This work is organized as follows. In Section \ref{sec:spin_flips} we outline the calculation of magnetic spectra and spin-flip rates and give an overview on the length scales and effects that have an impact on these quantities in a system with nonlocality.
Then, a specific nonlocal model for the bulk and surface response is introduced and used to obtain the near field asymptotes. We also briefly consider materials other than metals.
In Sec.\,\ref{sec:coherence} field correlation functions above
local and nonlocal metals are analyzed with respect to their
correlation length.
Sec.\,\ref{sec:discussion} summarizes the main results and reflects on their relevance for experimental setups.
Two Appendices give details on the electromagnetic Green's tensor near a surface and the calculation of the nonlocal reflection coefficients.

\section{Magnetic noise and spin-flip losses}
\label{sec:spin_flips}
\subsection{Noise spectra and spin flips}
\label{sec:noise_spectra}
%
%

Our main quantity of interest is $S^{B}_{ij}(\vec{r}, \omega)$, 
the spectral density per unit frequency
of the magnetic field cross-correlation ($i,j = x, y, z$).
This spectrum can be calculated with Green's function techniques, 
as outlined in Appendix \ref{app:greens_tensor}.
According to the fluctuation-dissipation theorem,
\begin{eqnarray}
S^{B}_{ij}(\vec{r}, \omega) &=& 2 \hbar\bar n( \omega ) 
\Im \mathcal{H}_{ij} ( \vec{r}, \vec{r}, \omega )
~.
\label{eq:spectrum}
\end{eqnarray}
where $\bar n( \omega )$ is the Bose-Einstein distribution
and $\mathcal{H}_{ij}$ the magnetic Green's tensor defined in 
Eq.(\ref{eq:def-Hij}).

Recall that the Green's tensor gives the field radiated by a pointlike
dipole source. 
In the presence of a surface, it therefore splits in two terms, the
first one being the same as in free space [Eq.\,(\ref{eq:GF_free_space_1point})], 
the second one describing
the magnetic field reflected from the surface. In the situations considered
here, the latter term dominates
the spectrum \cite{Purcell1946}.
For example, at a frequency of $100 \unit{MHz}$ and at room temperature, the spectrum of a surface at a distance of $1\unit{\mu m}$ exceeds the black body spectrum by 15 orders of magnitude.
The free space term in $\mathcal{H}$ can therefore be safely neglected and it is sufficient to consider the reflected Green's tensor, which is conveniently expressed in the Weyl representation as a two-dimensional Fourier integral
\begin{eqnarray}
\mathcal{H}_{ij} (z, \omega)	&= &\frac{\mu_0}{8 \pi } \int_0^\infty  dp\, p\kappa \left[ 
				\left(r_{\rm s}(\omega, p) 
				+ \frac{\omega^2}{c^2 \kappa^2} r_{\rm p}(\omega, p) \right) \times \right.\nonumber\\
				&&\left.\times [\delta_{ij} -  \hat{z}_i \hat{z}_j] 
				+ 2 \frac{p^2}{\kappa^2} r_{\rm s}(\omega, p) \hat{z}_i \hat{z}_j
				\right] e^{-2 \kappa z}~. 
\label{eq:mag_greentensor}
\end{eqnarray}
Here, s and p label the two principal polarizations,
$\kappa^2 = p^2 - \omega^2 / c^2$ is the propagation constant in 
vacuum, and $\hat{\vec{z}}$ the unit normal to the surface.
Details on the reflection coefficients $r_{\rm s}$, $r_{\rm p}$ 
are given in Appendix\,\ref{app:greens_tensor}.


Most of this work will consider near field noise, where large values of the 
perpendicular wave vector $p \gg \omega / c$ 
(evanescent waves) dominate the response and nonlocal effects become 
relevant.
In this regime, the p-polarization involving $r_{\rm p}$ is suppressed by the 
prefactor $\omega^2 / (c \kappa)^2$ in the integrand 
of Eq.\,\eqref{eq:mag_greentensor}. The analysis can thus be restricted to 
s-polarization for our purposes (magnetic field vector in the plane of incidence).

The magnetic noise spectrum has been measured via the loss rate of atoms from modern chip-based atom traps~\cite{Folman2002,Henkel2003, Fortagh2007, 
Reichel2011}.
An expression for the atomic transition rate due to fluctuations of the magnetic field can be obtained from Fermi's Golden Rule~\cite{Henkel2003} or 
a master equation approach~\cite{Henkel1999, Rekdal2004} 
\begin{eqnarray}
\Gamma_{a\to b}(\vec{r}) &=& \sum_{i,j} \frac{\mu_i^{ab} \mu_j^{ba}}{\hbar^2} S^{B}_{ij} (\vec{r}, -\omega_{ab}) 
	\label{eq:transition_rate}.
\end{eqnarray}
Here $a$ ($b$) labels a magnetic sublevel that is trapped (not trapped)
in the static magnetic field of the atom chip, $\mu_i^{ab} = \langle a |
\mu_i | b \rangle$ is the matrix element of the magnetic dipole operator,
and $\omega_{ab}$ the resonant Bohr frequency.
For magnetic moments in the order of a Bohr magneton $\mu_B$, the prefactor in Eq.\,\eqref{eq:transition_rate} translates a spectrum of {\nolinebreak  $130 \unit{pT^2 / Hz}$} to a transition rate of one per second. 
Rates in this low range have been measured near conducting surfaces
using ultracold atoms as a probe \cite{Jones2003,Harber2003,Lin2004}.
The magnetic near-field noise flips the spin of trapped atoms, leading to
loss from a magnetic trap and setting a fundamental limit to the coherence 
in these setups \cite{Folman2002, Henkel2003}. Conversely, this process
offers a way of probing material properties.




For atoms trapped in their electronic ground state, the magnetic moment
is dominated by the contribution of the electron spin. We evaluate the matrix 
elements
in Eq.\,(\ref{eq:transition_rate}) for simplicity by ignoring the quantum numbers
of the nuclear spin. We can then consider a two-level system of which one
state is magnetically trapped. 
For a static trapping field in the $xz$-plane that is tilted by an angle $\theta$ 
relative to the surface normal $\hat{\vec{z}}$, the  
magnetic dipole matrix elements read \cite{Henkel1999,Haakh2009b}
\begin{equation}
\boldsymbol{\mu}^{ge} = 
\frac{ \mu_B g_S }{ 2 }
\left( \cos \theta,
- {\rm i},
\sin \theta
\right)^T
.
	\label{eq:transition-dipole}
\end{equation}
Above a planar surface, the noise correlations are diagonal [see
Eq.(\ref{eq:lossrate_drude_scaling}) below] so that the spin-flip 
rate is proportional to $2 + \sin^2\theta$.
The more general case including hyperfine
structure can be found in Ref.\,\cite{Henkel1999}. 

\subsection{Overview: near-field noise}
\label{sec:overview}

\begin{figure}
a) \includegraphics*[width=7cm]{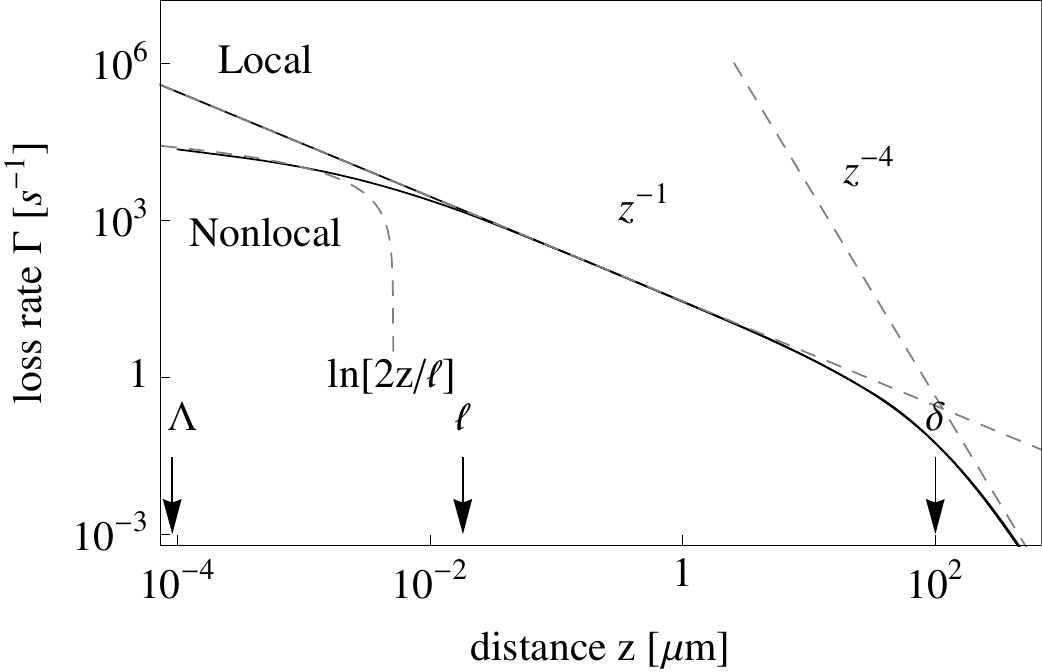}
\hspace{.5cm}\\
b)
\includegraphics*[width=7cm]{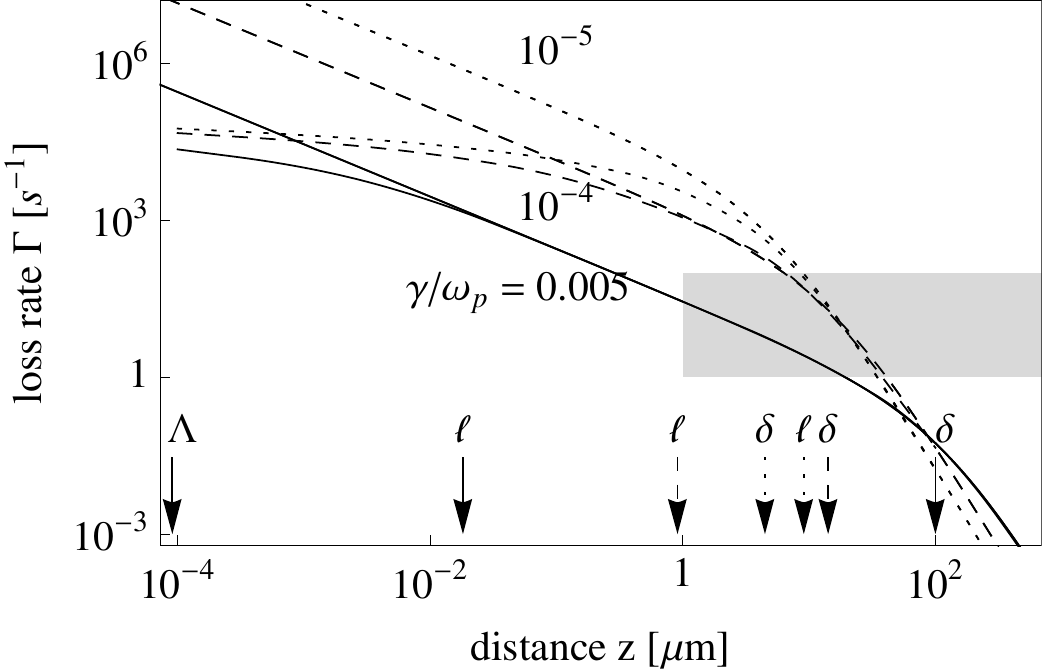}\\
\caption{a) Spin flip (loss) rate near a conducting half-space described by 
the nonlocal Boltzmann-Mermin conductivity~(\ref{eq:Boltzmann-Mermin1}--\ref{eq:Boltzmann-Mermin4}). The local description 
[Drude conductivity~(\ref{eq:Drude_model})]
and the asymptotic expressions of Eq.\,(\ref{eq:lossrate_drude_scaling})
(dashed) are shown for comparison. The length scales
$\Lambda = v_F / \omega_p$, $\ell = v_F / \gamma$, and
$\delta$ [Eq.\,(\ref{eq:def_skindepth})] illustrate the Thomas-Fermi 
screening length, the mean free path and the skin depth.
The parameters are for gold at $T = 300 \unit{K}$
($\sigma = 2.5 \times 10^7 \unit{\Omega^{-1} m^{-1}}$, $\ell = 18 \unit{nm}$, $\gamma = 6.7\times 10^{13}\unit{s^{-1}}$)
and 
the surface impedance is calculated with a specular boundary condition 
[Eqs.\,(\ref{eq:impedance_specular_s}, \ref{eq:impedance_specular_p})].
Spin flips are driven by fields at the Larmor frequency
$\omega / 2 \pi = 1 \unit{MHz}$ and oriented
parallel to the surface [$\theta = 0$
in Eq.\,(\ref{eq:transition-dipole})].
Losses due to the free space black body spectrum are much smaller
and not visible on this scale. 
\newline
b) Loss rates near gold surfaces with different purities. We vary the
ratio $\gamma/\omega_p$ between
relaxation rate and plasma frequency in the conductivity. The lowest curve
coincides with Fig.\,\ref{fig:lossrate_spec}a).  Note how
the intermediate regime $\ell \ll z \ll \delta$ opens up in the dirty limit.
The leftmost arrow marks the Thomas-Fermi screening length
$\Lambda = v_F / \omega_p$, while $\delta$ is the normal skin
depth~(\ref{eq:def_skindepth}).
}
\label{fig:lossrate_spec}
\end{figure}

Nonlocal effects can be expected to become visible on a length scale in the order of the mean free path $\ell$ of ballistic transport of charge carriers, as was already conjectured by Rytov and coworkers \cite{Rytov1989}.
Numerical calculations of the spin-flip rates for neutral atoms near a 
metal surface with and without a nonlocal response are shown in Fig.\,\ref{fig:lossrate_spec}a) and b).
Clear\-ly, there are three different asymptotic regimes of the distance between the atom and the surface, two of which involve distances much larger than $\ell$, where the surface spectrum cannot be distinguished from a local one. 
We shall find that in these regimes, the Green's tensor can be approximated by 
the scaling laws
%
\begin{eqnarray}
&& {\rm Im}\,\mathcal{H}_{ij} (\omega, z) =
\nonumber\\
&& \left\{ 
\begin{array}{c  l} 	\displaystyle
\frac{3 \mu_0 \delta(\omega)}{64 \pi  z^4 } [\delta_{ij} + 
\hat{z}_i \hat{z}_j], & \delta(\omega) \ll z
\\[1em]
	\displaystyle				
    \frac{\mu_0 }{32 \pi \delta^2(\omega) z} [\delta_{ij} + 
\hat{z}_i \hat{z}_j], 	&\ell \ll z \ll \delta(\omega)
\\[1em]
	\displaystyle
	\frac{ \mu_0 [\delta_{ij} + 
	\hat{z}_i \hat{z}_j]
	}{ 8\pi \delta^2(\omega ) \ell }
	\left( \ln\left[\frac{ \ell }{ 2 z } \right]- 0.077 \right)
	,
	&
	z \ll \ell~.
\end{array}
\right.\label{eq:lossrate_drude_scaling}
 \end{eqnarray}
Here, the skin depth of the normal skin effect is given by
\begin{equation}\label{eq:def_skindepth}
\delta(\omega) =  \sqrt{\frac{ 2 }{ \mu_0 \sigma \omega }}~,
\end{equation}
where $\sigma$ is the local limit of the DC conductivity [Eq.\,(\ref{eq:Drude_model}) 
below].
The local regime [first two lines in Eq.(\ref{eq:lossrate_drude_scaling})] 
were given already in Refs.\,\cite{Henkel1999a, Henkel1999}.

To make this qualitative behavior understandable, we propose an interpretation in terms of an active surface volume:

i) When $z \gg \delta(\omega)$, the normal skin effect screens noise 
from deep in the bulk so that only a skin layer of thickness $\delta( \omega )$
contributes to the noise. The noise is proportional to the squared non-retarded
fields $\sim 1/r^3$ of current loops , integrated over the surface -- 
this explains the power law $1/z^4$ and the proportionality to the
skin depth $\delta( \omega )$ in Eq.\,(\ref{eq:lossrate_drude_scaling}), upper
line.

ii) At smaller distances $\ell \ll z \ll \delta(\omega)$, a medium-filled
half-sphere of radius $\sim z$ effectively contributes to the noise. 
In addition, the probe particle now resolves individual current elements rather
than loops. The noise then arises from the squared fields of these current 
elements ($\sim 1/r^2$), integrated over the volume
of the half-sphere, as explained in Ref.\,\cite{Henkel2005c}. 


The previous cases i) and ii) have been observed experimentally
in the kHz to MHz range
with sensitive magnetometers \cite{Varpula1984} and with trapped ultracold
atoms \cite{Harber2003}. In this paper, we address the regime

iii) of the extreme near field in a nonlocal conductor, $z \ll \ell$: 
the ballistic (rather than diffuse) motion of charge carriers creates 
spatial correlations in the fluctuating current field. This reduces 
the number of mutually uncorrelated volume elements in the half-sphere
introduced in ii) above, and hence lowers the noise power. Note that the limiting 
value in Eq.\,(\ref{eq:lossrate_drude_scaling}) scales with $1/(\delta^2\ell)$
which is actually independent
of the relaxation time in a Drude conductor. The magnetic noise is related
to Landau damping, or equivalently to the thermal excitation of electron-hole 
pairs \cite{Ford1984}. This regime is therefore quite universal, and
we show in Secs.~\ref{sec:semiconductors} and \ref{sec:superconductors} that semiconductors and even 
superconductors follow the same scaling law.

The rest of this section will review the nonlocal response functions of the
conductor and present calculations that confirm these arguments.

\subsection{Model of a nonlocal metal}
\label{sec:nonlocal_model}

%

In the nonlocal regime, the current-field relationship of the bulk, i.e. the conductivity or the dielectric tensor, depends on the wave vector and it is necessary to 
distinguish longitudinal and transverse response functions.

\begin{figure}
a)
\includegraphics[height=5cm]{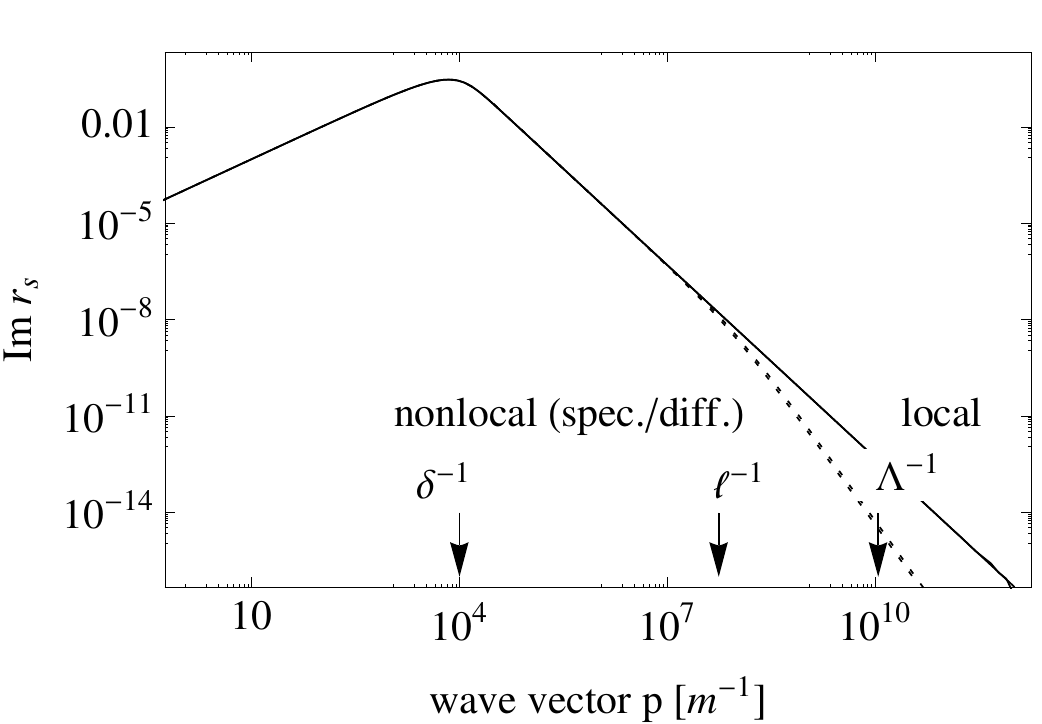}\hspace{.6cm}
\\[1em]
b)
\includegraphics[height=5cm]{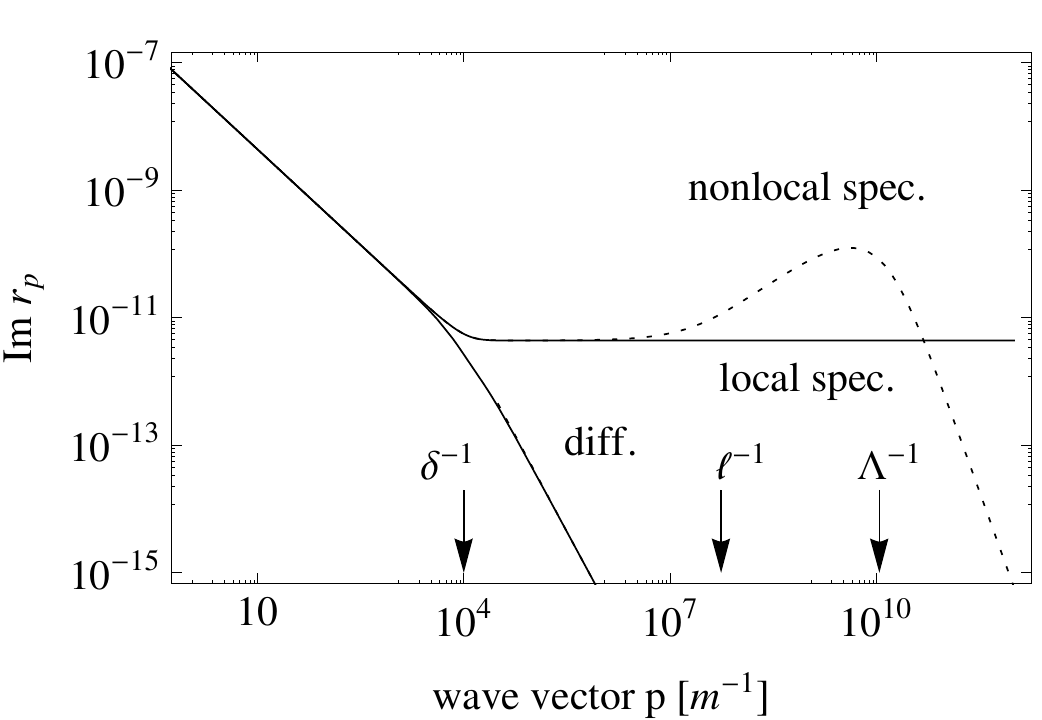}
\caption{Reflection coefficients $\Im r_{s,p}$ vs. wave vector in the evanescent sector. a) s-polarization and b) p-polarization.
The curves show specular and diffuse scattering at the boundary, and bulk responses given by the local Drude model (solid), and in the nonlocal Boltzmann-Mermin model (dotted). Parameters are for gold
($\sigma = 2.5 \times 10^7 \unit{\Omega^{-1} m^{-1}}$, $\ell = 18 \unit{nm}$, $\gamma = 6.7\times 10^{13}\unit{s^{-1}}$).
Frequency $\omega /2\pi = 1\,\unit{MHz}$.}
\label{fig:reflectivity_specular_MHz}
\end{figure}

Very basic descriptions taking into account some nonlocal effects 
are hydrodynamic models, see, e.g., Ref.\cite{Flores1979, Barton1979}. 
These approaches are valid in a restricted momentum range.  The simplest
model does not lead to any 
change in the s-polarization with respect to the local limit, unless some
some phenomenological transverse response is introduced.
A more substantial description of a metal is given by 
the Boltz\-mann-Mer\-min (BM) model \cite{Lindhard1954,Ford1984, Mermin1970}
\begin{align}
\sigma_l(\omega, \vec{k}) &= 
- \frac{\gamma\,\sigma 
	}{\gamma - {\rm i} \omega}
	\frac{3 \omega \, u^2 f_l(0,u)}{\omega + i \gamma f_l(0,u)}
	\label{eq:Boltzmann-Mermin1}
\\
\sigma_t(\omega,\vec{k}) &= 
\frac{\gamma\, \sigma 
	}{ \gamma - {\rm i}\omega } 
	f_t(0,u)
~,
\label{eq:Boltzmann-Mermin2}
\end{align}
where $\sigma$ is the DC conductivity in the local limit, $\gamma$ describes
the broadening of the electronic states at the Fermi level
due to scattering. The variable $u$ contains the
momentum dependence via the dimensionless functions
\begin{align}
f_l(0,u) &= 1- \frac{u}{2} \ln \left[\frac{u+1}{u-1}\right], \quad u = \frac{\omega + i \gamma}{k v_F}
\label{eq:Boltzmann-Mermin3}\\
f_t(0,u) &= \frac{3}{2} u^2 - \frac{3}{4}u (u^2-1) \ln \left[\frac{u+1}{u-1}\right]
~,
\label{eq:Boltzmann-Mermin4}
\end{align}
where $v_F$ is the Fermi velocity. and
the logarithm is taken with a branch cut along the negative real axis.
These expressions are obtained in the limit $k/2 k_F \ll 1$ from a more general model due to Lindhard\cite{Ford1984},
hence the redundant first argument 0.
This assumption is reasonable because for our purposes, the relevant wave
vectors are in the range $k \sim 1/z$, much smaller than the Fermi
momentum $k_F \approx 10^{10} \unit{m^{-1}}$.

The relevant frequencies for magnetic transitions, typically in the rf- to 
microwave range, 
lie in the Ha\-gen-Ru\-bens regime $\omega \ll \gamma \ll \omega_p$ where $\omega_p$ is the conductor's
plasma frequency.
In this case, spatial dispersion is obviously encoded by the parameter
$u \approx 
{\rm i} / k \ell$, and 
the mean free path $\ell = v_F / \gamma$ sets the relevant scale.
In the local limit $k \ell \ll 1$, both conductivities reduce to the
Drude form 
%
%
%
\begin{align}\label{eq:Drude_model}
\sigma_{l,t}(\omega, \vec{k})  \to
\sigma(\omega) &= \frac{\omega_p^2 \varepsilon_0}{ \gamma - {\rm i} \omega }
~.
\end{align}
This is the regime of the normal skin effect where Eq.\,\eqref{eq:def_skindepth} applies.

While the conductivity describes the bulk response of the conductor, the specific properties of the surface have an impact on the nonlocality of 
response, too.
The calculation of the reflectivities requires the solution of the electromagnetic scattering problem at the surface. If the bulk conductivity depends on the wave vector, Fresnel's equations do not hold any more, and one has to introduce additional boundary conditions for the current density at the inner surface.
The latter are modeling the way charge carriers are scattered there.

The simplest assumption is that of specular reflection of charge 
carriers\,\cite{Ford1984,Reuter1948,Kliewer1968,Flores1979}.
Diffuse scattering \cite{Flores1979, Kliewer1970, Foley1975} or a general 
combination of both mechanisms can be included, but must be treated with
care to ensure that charge conservation holds at the surface \cite{Flores1979}.
Severe as it may be, this problem only occurs when electric fields have components perpendicular to the surface, i.e. in the p-polarization, while
the nonlocal effects considered in this work involve the s-polarization.
In addition, it is well known that the scattering mechanisms give little differences 
for the anomalous skin effect \cite{Reuter1948}. Much larger corrections occur, e.g. due to surface roughness \cite{BedeauxVlieger}. 

The calculation of reflection coefficients at the surface of a nonlocal metal is described in Appendix\,\ref{app:reflectivity}. The resulting reflection amplitudes
are shown in Fig.\,\ref{fig:reflectivity_specular_MHz} for both polarizations.
We plot the absorption ${\rm Im}\,r_{s,p}$ which is proportional, by reciprocity, to 
the radiated noise power.
In the s-polarization, all models converge to the local scenario as $p \ll 1/\ell$,
as expected. Spatial dispersion leads to reduced noise 
for wave vectors $p \gg 1/\ell$. 
The impact of nonlocality is much more important in p-polarization.
The increase of p-polarized absorption
in the range $1/\ell < p < 1/\Lambda$ has been discussed previously
\cite{Ford1984, Larkin2004, Henkel2006, Chapuis2008}; it is due to the internal 
photo-effect 
(creation of particle-hole pairs, Landau damping). The results with the
diffuse boundary condition introduced in Ref.\,\cite{Foley1975} deviate 
from the local limit already in the range $p \gtrsim 1/\delta$, and 
become independent of the bulk conductivity.
This is likely to be an artifact 
due to the violation of charge conservation, as discussed in
Refs.\,\cite{Flores1977, Flores1979, Foley1975}.

Note that the factor
$(\omega / c \kappa)^2 \sim 10^{-16}$ is very small
where the p-polarized
absorption peaks so that it is a good approximation to neglect this part 
in the Green's function~(\ref{eq:mag_greentensor}). This polarization
is essential, on the contrary, 
for situations sensitive to surface charges and electric surface 
fields~\cite{Flores1979}, 
such as heat transport~\cite{Volokitin2007}, heating of trapped ions 
\cite{Henkel1999} or the electric dipole contribution 
to dispersion forces~\cite{ParsegianBook}.%

\subsection{Short-distance approximation}


We derive here the asymptotic form of Eq.\,(\ref{eq:lossrate_drude_scaling}),
third line. Within the approximations introduced above, the Green 
tensor~(\ref{eq:mag_greentensor}) can be calculated from
\begin{eqnarray}
	&&
	\mathcal{H}_{ij}( z, \omega ) \approx
	\frac{ \mu_0 [\delta_{ij} + \hat{z}_i \hat{z}_j] }{ 8\pi }
	\int\limits_0^{\infty}\!{\rm d}p \, p^2 
	r_{\rm s}( \omega, p) \, {\rm e}^{ - 2 p z }~.
	\label{eq:approximate-integral-for-Hij}
\end{eqnarray}
The distance range is now $z \gg \ell$ so that 
the relevant wave vector range is $p \sim 1/z \gg 1/\ell$.
We start from an expansion of the Boltzmann-Mermin 
model~\eqref{eq:Boltzmann-Mermin1}-\eqref{eq:Boltzmann-Mermin2} 
at small values of $u$.
The limiting form of the reflection coefficients for specular scattering
of charge carriers is found as (see Appendix\, \ref{app:reflectivity})
\begin{eqnarray}
p \gg 1/\ell, \mbox{ spec. scatt.:} \quad
\Im r_{\rm s}( p, \omega )
			&\approx& 
			\frac{1}{p^3 \delta^2( \omega ) \ell} 
			\label{eq:rs_specular_BM_large_k}
\end{eqnarray} 
while the diffuse boundary condition yields a result smaller by a factor $3/4$.
This is similar to the findings of Reuter and Sondheimer \cite{Reuter1948} for
the anomalous skin effect.
The following analysis assumes specular scattering.  
The power law of Eq.\,(\ref{eq:rs_specular_BM_large_k}) illustrates the 
reduction of noise by spatial dispersion (the decay with momentum $p$ 
is faster) and agrees well with
a numerical calculation, as illustrated in Fig.\,\ref{fig:reflectivity_specular_MHz}. 

%
%
We split the integration range at $p \sim 1/\ell$ and replace for
$p \gg 1/\ell$ the reflection coefficient $r_{\rm s}( \omega, p )$ by the
nonlocal approximation~(\ref{eq:rs_specular_BM_large_k}). 
The integral then gives 
\begin{eqnarray}
	&& \int\limits_{1/\ell}^{\infty}\!{\rm d}p \, p^2 
	r_{\rm s}( \omega, p) \, {\rm e}^{ - 2 p z }
	\approx
	\frac{ 1 }{ \delta^2( \omega ) \ell}
  	\int_{1/\ell}^\infty \frac{ {\rm d}p }{ p }  e^{-2 p z}
\nonumber\\
	&&=
	\frac{ \text{E}_1(2 z / \ell) }{ \delta^2( \omega ) \ell}
	\approx
	\frac{ \ln(\ell/2 z) - \gamma_E }{ \delta^2( \omega ) \ell }~.
\label{eq:greentensor_nearfield_specular} 
\end{eqnarray}
Here,
$\text{E}_1(z) = \int_{z}^{\infty }\!{\rm d}t\, t^{-1} e^{- z t}$ is the 
exponential integral and 
$\gamma_E \approx 0.577$ is the Euler-Mascheroni constant.
In the range 
$p \ll 1/\ell$, the reflection coefficient $r_{\rm s}( \omega, p )$ is approximately
equal to its local 
form [Eqs.(\ref{eq:reflectivity_impedance}, \ref{eq:impedance_fresnel})]. 
The integral then gives (we assume $\ell \ll \delta$)
\begin{equation}
 	\int_0^{1/\ell} p^2 \, dp \Im r_{\rm s}(\omega, p)e^{-2 p z}
	\approx \frac{ 1 + {\cal O}( \ell / \delta ) }{ 2 \delta^2( \omega ) \ell}~. 
	\label{eq:contribution-from-local-range}
\end{equation}
Summing the two contributions gives the approximate Green's tensor
(always for $z \ll \ell$)
\begin{equation}
	{\rm Im}\,\mathcal{H}_{ij}( z, \omega ) \approx
	\frac{ \mu_0 [\delta_{ij} + \hat{z}_i \hat{z}_j] }{ 
		8\pi \delta^2( \omega ) \ell }
	\left(
	\ln\left[\frac{\ell}{2 z}\right] 
	- \gamma_E + \frac12
	\right)~.
	\label{eq:eq:greentensor_nearfield_specular_ansatz}
\end{equation}
%
%
%
This is the third regime of the Green's tensor~(\ref{eq:lossrate_drude_scaling})
discussed in Sec.\,\ref{sec:overview} and corroborates the statement
that a nonlocal description predicts less noise at short distances compared
to a local one.
We have thus generalized a similar result reported in Ref.\,\cite{Chklovskii1992} 
within the context of nuclear spin relaxation, where the normal skin effect was 
neglected. 
We conclude that for the miniaturization of atom chip experiments, a large 
mean free path is advantageous. Crystalline metals may push $\ell$ into
a range that is achievable with atom chip traps.
The other possibility may be chips based on pure semiconductor substrates
that we discuss now.



\subsection{Semiconductors}
\label{sec:semiconductors}
The previous analysis can be generalized to other classes of conducting materials.
Strongly doped semiconductors (where the electron gas is degenerate like
in a metal) may be described by the BM model \eqref{eq:Boltzmann-Mermin1}-\eqref{eq:Boltzmann-Mermin4} with modifications only in the values of 
the parameters. 
For frequencies well below the gap, a background dielectric constant 
$\varepsilon_b$ appears due to the static interband polarizability,
but this does not play a role for magnetic near field noise.
Weakly doped semiconductors (non-degenerate electron gas), on the other hand, are not ruled by Fermi statistics, but by a thermal distribution with the characteristic velocity $v_T = (3 k_B T / m)^{1/2}$ taking over the role of 
the Fermi velocity $v_F$.
Here, Eqs.\,\eqref{eq:Boltzmann-Mermin3} and~%
\eqref{eq:Boltzmann-Mermin4} should be replaced by\,\cite{Landau1981,Melrose1991}
\begin{align}
f_t(0,u) &= 1 - \frac{3}{2} f_l (0,u) = - \frac{u}{\sqrt{\pi}}\int_{-\infty}^\infty dz \frac{\exp(-z^2)}{z-u}
~.
\end{align}
We observe that this is numerically very close to Eqs.\,\eqref{eq:Boltzmann-Mermin3}, \eqref{eq:Boltzmann-Mermin4} in the local regime
$k \ell \ll 1$ and differs only by a numerical factor in the deeply nonlocal 
regime. 
Up to this changed prefactor, the preceding calculations for the 
$r_{\rm s}$ coefficient and the Green's tensor carry through so that the physics
is qualitatively the same.

Let us consider typical numbers that can be found from experiments 
on charge transport in silicon \cite{Weber1991}. An n-type semiconductor with a rather low doping of $3\times 10^{16} \unit{cm}^{-3}$ is characterized by 
a skin depth $\delta > 1 \unit {cm}$ in the $\unit{MHz}$ range.
At room temperature, $v_T = 1.2 \times 10^5\unit{m/s}$ and  
$\ell \approx 60 \unit{nm}$, quite comparable to the value for gold.
Since it is proportional to the DC conductivity, the 
magnetic near field spectrum is smaller 
by orders of magnitude compared to a metal.
Cooling the sample down to $100 \unit{K}$ reduces phonon excitations and
enhances the conductivity. Yet, the thermal velocity drops also to 
$v_T = 6.7 \times 10^4\unit{m/s}$, so that the mean free path is barely
larger, $\ell \approx 85\unit{nm}$.
At even lower temperatures the conduction band occupation freezes out 
and the response becomes local.

\subsection{Superconductors}
\label{sec:superconductors}

\begin{figure}
\includegraphics[width=7cm]{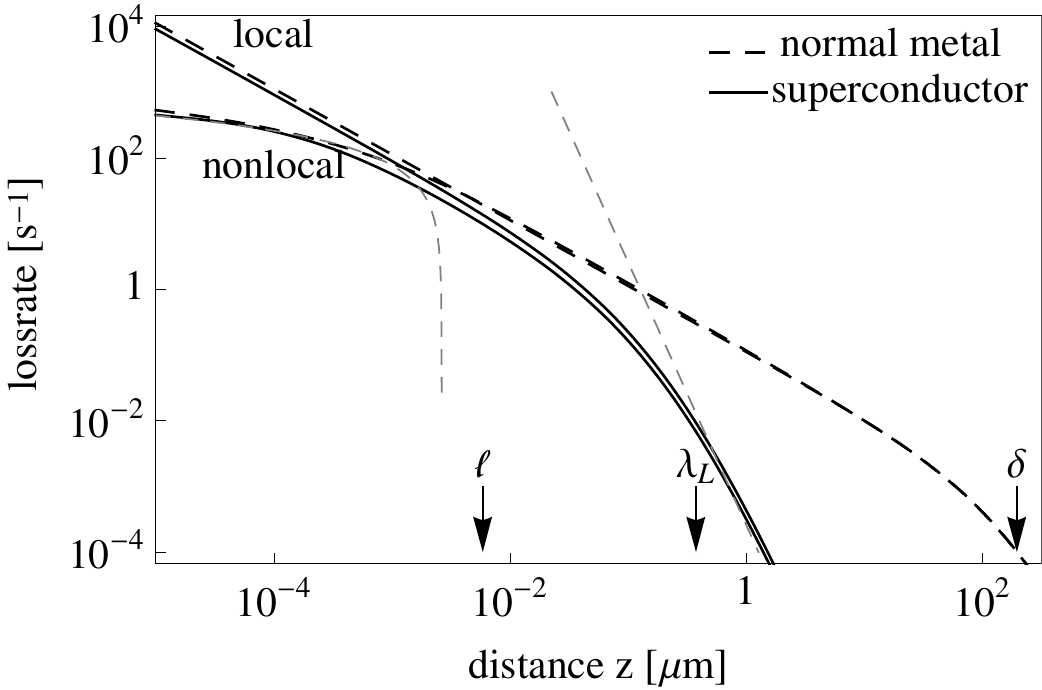}
\caption{
Spin flip (loss) rate near a surface made of superconducting niobium 
(solid curves) and a fictitious normal metal (dashed curves). Nonlocal effects are given for both cases as compared to the local limit.
Parameters for niobium  at $T = 0.5 T_c$ follow Ref.\,\cite{Ashcroft1987}
$\omega_p = 1.33 \times 10^{16} \unit{s^{-1}},
\gamma = 2.38\times 10^{-14} \unit{s^{-1}},
T_c = 9.2 \unit{K}, 
\Delta(0) = 1.9 \, k_B T_c =
h \times 7.9 \unit{GHz}$,
$\ell = 5.7\unit{nm}$.
We chose the
Larmor frequency $\omega / 2 \pi = 1 \unit{MHz}$ and  $\theta = 0$. Note that Ref.\,\cite{Popel1989} gives a mean free path larger by a factor of 4 and a smaller value of $\gamma$.
For better comparison the normal metal is obtained by closing the superconducting energy gap and rescaling the plasma frequency.
Gray dashed lines indicate the asymptotes to the superconductor.
Losses due to the free space black body spectrum are not visible on this scale.
}
\label{fig:lossrate_nlbcs}
\end{figure}


The nonlocal response of superconductors in the micro\-wave range was discussed in Refs.\,\cite{Popel1989, Miller1960}. It was argued by Rickayzen \cite{Rickayzen1959} that screening in a superconductor does not differ greatly from 
a normal metal. This is because all charge carriers contribute to screening, while the specific properties of a superconductor are determined by the states close 
to the Fermi level. We thus expect the noise spectrum to be characterized by the same logarithmic asymptote \eqref{eq:greentensor_nearfield_specular} found earlier. This is indeed confirmed by numerical calculations of magnetic noise near a niobium surface, the results of which are
shown in Fig. \ref{fig:lossrate_nlbcs}.
We have evaluated the non-local BCS conductivity within the approach of
Mattis and Bardeen including disorder scattering \cite{Mattis1958}, using the
expressions of P\"opel~\cite{Popel1989}. The local limit recovers correctly the results from Refs.\,\cite{Zimmermann1991, Berlinsky1993}. For comparison we also give the curves for a fictitious normal metal, with the same parameters
except that the gap is closed ($\Delta(0) = 0$). The description then
coincides with the nonlocal BM model. The metallic plasma frequency
was adjusted in order to take into account the redistribution of the spectral
weight by disorder, 
as discussed in Refs.\,\cite{Berlinsky1993, Bimonte2010}.

In the local regime $z \gg \ell$, the superconductor indeed shows a strong
reduction of magnetic noise. This happens because the relevant frequency
is below the gap, $\hbar\omega \ll \Delta( T)$, and magnetic fields are well
screened by the Meissner effect.
At large distances, we find good agreement with the expressions for 
the lossrate obtained by Skagerstam et al. \cite{Skagerstam2006} for a 
two-fluid model, if the same re-scaled plasma frequency is taken into account 
(gray dashed asymptote in Fig.\,\ref{fig:lossrate_nlbcs}). In terms of the
Green's tensor,
%
\begin{equation}
\Im \mathcal{H}_{xx}(z, \omega) \approx \frac{\mu_0 \lambda_L^3(T)}{4 \pi \delta^2(\omega, T) z^4} \left(\frac{3}{4}\right)^2~.\end{equation}
The Meissner-London length
$\lambda_L( T) = (c / \omega_p ) \Delta(0)/ \Delta(T) \approx (c / \omega_p)  
[1 - (T/T_c)^4]^{-1}$
determines the penetration depth for quasistatic fields, while the skin
depth $\delta(\omega, T)$ involves the conductivity for the normal fluid fraction.
The Meissner effect becomes inefficient, however, if the spatial scale $1/p$ of 
the noise field becomes comparable or smaller than the penetration depth 
$\lambda_L( T )$. The loss rate for $z \ll \lambda_L( T )$ then approaches 
the $1/z$ asymptote of a normal conductor, see Fig.~\ref{fig:lossrate_nlbcs}.
At shorter length scales $z \ll \ell$, we recover the logarithmic 
scaling law found before for the normal conductor (thin dashed line). 
This illustrates the 
very general character of this regime that does not depend greatly on the
material class. 

\section{Lateral coherence}
\label{sec:coherence}
A nonlocal conductivity creates spatial correlations in the current fluctuations 
below the
surface which reduce the overall magnetic noise level. It is to
be expected that this leaves also a signature in the correlations of the
field. These correlations are universal for blackbody 
radiation~\cite{MandelWolf} and have been studied in 
Refs.~\cite{Gori1994, Wolf2001, Blomstedt2007} for homogeneous media
and in Refs.~\cite{Carminati1999, Henkel2000, Dorofeyev2002, Henkel2006,
Lau2007, Norrman2011} 
for the near field 
of bodies. The spatial correlation length can be much larger or much smaller
than the wavelength, depending on the polariton modes that dominate the
electromagnetic field noise.
We find in this section that the correlation length is connected to the mean 
free path as a direct consequence of the ballistic motion of the charge carriers.
This should be contrasted to electric fields near nonlocal solids where
the spatial correlations were found to differ from the local description only
at distances comparable to the Thomas-Fermi length $\Lambda$, with
the mean free path $\ell$ not playing any role \cite{Henkel2006}.

We are interested in the correlation between fields
at a fixed height $z$ from the surface and laterally separated by 
a distance $\rho$. 
We define the coherence function as the cross-correlation spectrum
of the normally ordered field operators in frequency space: 
\begin{eqnarray}
 \mathcal{B}_{ij}(\rho,z, \omega) 
	&=& \int d t \langle : B_i(\rho,z,t) B_j(0,z, 0) : \rangle 
	{\rm e}^{{\rm i} \omega t}~.
\end{eqnarray}
The
fluctuation-dissipation theorem for normally ordered operator products \cite{Agarwal1975} provides the link to the two-point Green's tensor
\begin{eqnarray}
\mathcal{B}_{ij}(\rho, z, \omega) = 2 \hbar \bar n
\Im \mathcal{H}_{ij}(\vec{r}, \vec{r}', \omega)~,
\end{eqnarray}
where the points $\vec{r}$, $\vec{r}'$ are located at the same height and
laterally separated by $\rho$. This generalizes Eq.\,\eqref{eq:spectrum}. 
A general integral form
is given in Appendix~\ref{app:greens_tensor}. We drop the frequency
arguments for simplicity in the following.

\subsection{Local limit}
Where a local description of the metal is sufficient, the coherence function in the near field can be evaluated asymptotically by expanding all integrands for values 
$p \gg 1/\delta, \omega / c $. The resulting integrals have the form $\int_0^\infty dp\, e^{-2 p z} J_{n}(p \rho) p^m$, ($n, m \in \mathbbm{N}_0$) and can be evaluated exactly.

\begin{figure}[h!!]
\centering
a) \includegraphics[width=7cm]{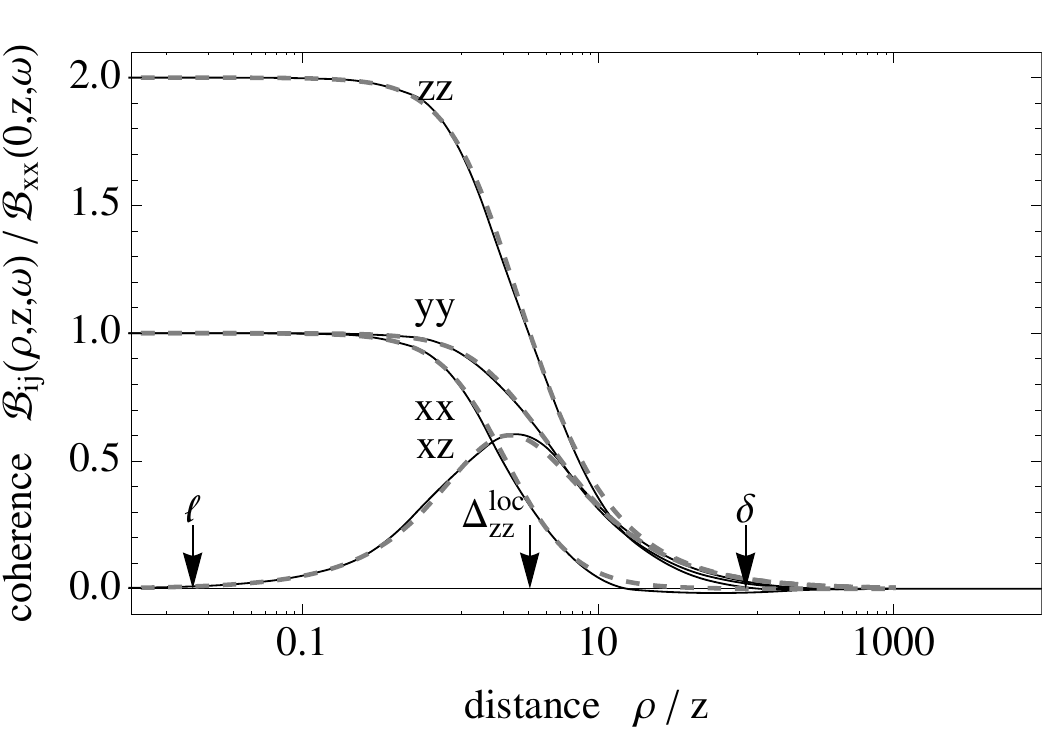}\\
b) \includegraphics[width=7cm]{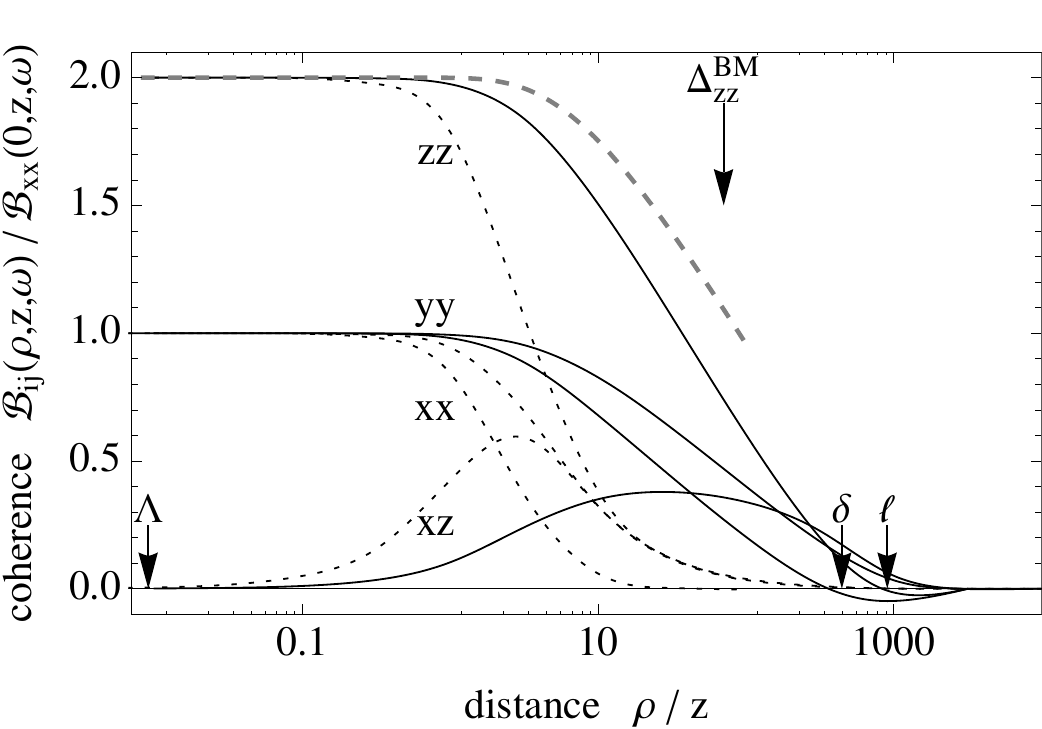}
\caption{%
Components of the coherence tensor of the magnetic field 
near a gold surface.
Solid (dotted) lines: Boltzmann-Mermin model with specular boundary
condition (Drude model).
a) Distance $z = 1 \unit{\mu m}$ much larger than mean free path
$\ell = 18\,\unit{nm}$: all curves coincide 
with the local approximation 
(\ref{eq:tensor-display-lateral-correlations})
(dashed gray curves).
b) Nonlocality becomes important at short distances.
For better visibility we set $\ell = 10\unit{\mu m} \gg
z = 10 \unit{nm}$ by lowering the scattering rate $\gamma$.
The thick dashed line gives the rather coarse approximation from Eq.\,\eqref{eq:coherence_mag_nl_zz_2}  (dashed gray curve).
Frequency $\omega/2\pi = 1\,\unit{MHz}$. 
}
\label{fig:coherence}
\end{figure}

We find that at distances $\ell \ll z \ll \delta$, the magnetic 
coherence tensor depends on the distance $d = \sqrt{ \rho^2 + (2z)^2 }$
between one observation point and the mirror image of the other.
The tensor elements are very well approximated by
(see Fig.\,\ref{fig:coherence})
%
\begin{equation}
	\mathcal{B}_{ij}( \rho, z ) =
	\frac{ 4 z^2 \, 	\mathcal{B}_{xx}( 0, z ) 
	}{ d ( d + 2 z ) }
	\left(
	\begin{array}{ccc}
	2 &   & \rho / z
	\\
	 & d/z & 
	\\
	\rho / z & & 2 + d/z
	\end{array}
	\right)~.
	\label{eq:tensor-display-lateral-correlations}
\end{equation}
Here, the noise spectrum $\mathcal{B}_{xx}(0,z)$ 
for $\rho = 0$
was introduced as a convenient scale [see Eq.\,(\ref{eq:lossrate_drude_scaling})].
The axes are chosen such that
the $x$-axis points along the separation between the two observation
points.
These quantities are independent of the specific material properties and depend only on the ratio $\rho / z$, i.e. the geometry of the system.
An equivalent form for the $xx$ component was already given 
in Eq.\,(33) of Ref.\,\cite{Henkel2003}, see also Ref.\cite{Nenonen1996}.
Note that the $xz$ cross-correlation was missed in Ref.\cite{Lau2007}.

The coherence functions decay on a typical length scale. For example, the $zz$-component (and similarly for the other ones) is characterized by the correlation length
\begin{equation}
\ell \ll z: \quad \Delta_{zz}^{\rm loc} = 2 \sqrt{3} \, z
~,
	\label{eq:coherence-length-local}
\end{equation}
where $\mathcal{B}_{zz}( \rho, z )$ drops to half its value at $\rho = 0$
(see Fig.\ref{fig:coherence}a)).
The approximate forms of Eq.\,(\ref{eq:tensor-display-lateral-correlations})
are not valid far beyond the correlation length
where some correlation functions become negative,
as shown in Fig.\,\ref{fig:coherence}a). The agreement with the asymptotes
is so high, however, that the curves are hardly distinguishable. The peak
in the crossed $xz$-correlation arises from light paths that are reflected
from the surface at oblique angles and whose fields are polarized in the 
$xz$-plane.

\subsection{Nonlocal metal}
In the near field $z \ll \ell$, nonlocality leads to a larger coherence length 
than predicted by the local scenario, as is clearly visible 
in Fig.\,\ref{fig:coherence}b).

In this section we extract the relevant scales by evaluating the two-point
Green's tensor $\mathcal{H}_{zz}( \rho, z )$. 
It is necessary to interpolate the reflection coefficient $r_{\rm s}( p )$ between 
the local and the nonlocal limits
to avoid an unphysical logarithmic divergence at the lower bound.
The simplest choice is the Pad\'e
approximation
\begin{equation}
	\Im r_{\rm s}( p ) \approx 
	\frac{\Im r_{\rm s}^{\rm loc}( p )}{1 + p \ell / 2} 
	\approx \frac{1}{2 \delta^2 p^2 (1 + p \ell / 2)}~,
	\label{eq:interpolation-rs}
\end{equation}
where the last form is appropriate for $p \gg 1/\delta$ and the small-$p$
divergence
is removed by the factor $p^2$ under the integral
[see, e.g. Eq.\,(\ref{eq:approximate-integral-for-Hij})].
The integrals then give 
%
\begin{eqnarray}
\mathcal{H}_{zz}(0, z) 
&\approx& 
	\frac{\mu_0}{4 \pi \delta^2\ell} \, \text{E}_1(4 z/ \ell) 
	{\rm e}^{4 z / \ell}
	~,
\\
\mathcal{H}_{zz}( \rho, z) 
&\approx& 
	\frac{\mu_0}{8 \pi \delta^2} 
	\int_0^{\infty} dp \, \frac{\exp(-2 p z) J_0(p \rho)}{1 + p \ell / 2}
~,
	\label{eq:extreme-near-field-asymptote-Hxx}
\end{eqnarray}
where $\text{E}_1$ is the exponential integral defined after 
Eq.\,(\ref{eq:greentensor_nearfield_specular}).
We split the integral at $p = x_0 / \rho$ where $x_0 \approx 2.356$ 
is the first zero of the Bessel function $J_0(x)$. For $0 < p < x_0 / \rho$, 
$J_0( p \rho )$ 
is replaced by a spline $j(x) = 1 + b x^2 + a x^3$,
and replaced 
by its asymptote $j(x) \approx \sqrt{2/\pi x} \cos(\pi/4 - x)$ for larger
arguments.
It turns out that the first interval gives the dominant contribution, 
since the oscillations beyond $x_0/\rho$ provide a cut-off for the integrand. 
We obtain 
%
\begin{eqnarray}
\mathcal{H}_{zz}(\rho, z) 
&\approx&
\mathcal{H}_{zz}(0,z) 
- \frac{\mu_0}{16 \pi \delta^2\ell}\biggl[
- 4\, \text{E}_1( 2 x_0 z / \rho ) 
	\nonumber\\
&&
+ \frac{b \rho^2}{ z^2} 
\left(
1 - (1 + 2 x_0 z/\rho) \, {\rm e}^{ - 2 x_0 z/\rho }
\right)
	\nonumber\\
&&
+ \frac{a \rho^3}{ z^3}
\left(
1 - (1 + 2 x_0 z/\rho + 2 (x_0 z / \rho )^2) 
	\, {\rm e}^{ - 2 x_0 z/\rho }
\right)
	\nonumber\\
&&
+ \mathcal{O}\left(\frac{\rho}{ \ell},\frac{z}{ \ell}\right)
\biggr] 
	\label{eq:coherence_mag_nl_zz}
~.
\end{eqnarray}
A careful glance at this expression shows that in the regime $z \ll \rho \ll \ell$
, the first line dominates. The decay of
the lateral coherence is therefore logarithmically
(see Fig.\,\ref{fig:coherence}), as
the small-argument approximation to the exponential integral
illustrates (thick dashed line in Fig.\ref{fig:coherence}b))
\begin{eqnarray}
\frac{\mathcal{B}^{\rm BM}_{zz}(\rho, z)}{\mathcal{B}^{\rm BM}_{xx}(0,z)} 
&\approx& 
2 - \frac{2\text{ E}_1( 2 x_0 z / \rho )}{ \text{E}_1( 4 z / \ell )}
	\label{eq:coherence_mag_nl_zz_2}
\approx 2 - \frac{\ln\left[(2 x_0 z / \rho)^2 \right]
	}{ \ln\left[4 z / \ell\right]}
	~.
\nonumber\\
\end{eqnarray}
%
%
%
%
%
%
This yields a coherence length
%
\begin{equation}
z \ll \ell: \quad
\Delta_{zz}^{\rm BM} =  x_0 \sqrt{ z \ell}
~
	\label{eq:coherence-length-non-local}
\end{equation}
much larger than its local counterpart~(\ref{eq:coherence-length-local}).
%
The spatial correlations of magnetic near fields are therefore 
linked to the characteristic mean free path of ballistic transport.
We recall that a similar discussion in Ref.\,\cite{Henkel2006} for electric
correlations did not find Eq.\,(\ref{eq:coherence-length-non-local}) involving
the mean free path $\ell$, but rather the Thomas-Fermi screening length 
$\Lambda$ which is typically smaller. The impact of spatial dispersion 
is therefore somewhat easier to reveal by analyzing magnetic fields.

\section{Discussion}
\label{sec:discussion}
We have found that the scattering mean free path of charge carriers sets the 
distance scale for the onset of nonlocal effects in the near field noise.
For evanescent modes that dominate the near field, the 
fraction of the metallic volume that contributes to noise is limited by 
screening more efficiently than by the normal skin effect.
The present calculation indicates, therefore, that loss rates 
are actually lower at short distances than 
predicted by local conductivity models (Ohm's law). This noise reduction
sets in at atom-surface separations comparable to or below the mean free 
path. 
In a clean (crystalline) metal $\ell$ may take values in the order 
of $1 \unit{\mu m}$ which is at the limits of the experimentally 
accessible region (cf. the gray box in Fig.\,\ref{fig:lossrate_spec}): 
typical traps operate at distances of $1 \ldots 100 \unit{\mu m}$ from 
the surface 
and can resolve lifetimes up to $10^{-2} \dots 1 \unit{s}$. 
The data shown in
Fig.\,\ref{fig:lossrate_spec} are calculated at room temperature.
Since both the conductivity and the mean free path depend on temperature,
it is worth investigating whether the lifetimes of magnetic levels
may be tuned by cooling the atom chip device.
This strategy is hitting a limit in the extreme near field (distance
$z \ll \ell$): the noise power becomes independent of the
scattering rate of carriers (the Drude parameter 
$\gamma$), and also the details of the 
scattering mechanism of charge carriers at the inner surface become
irrelevant. The 
near field spectrum [given in Eq.\,\eqref{eq:greentensor_nearfield_specular}] 
has a rather general character and is expected to apply to doped 
semiconductors and even to superconductors, as our numerical calculations 
show (Fig.\,\ref{fig:lossrate_nlbcs}). 

We have also analyzed the asymptotic form of the noise correlations in the 
short-distance range and found that 
magnetic fields are laterally coherent on a scale $\sim \sqrt{(\ell + z) z}$.
This implies for an atom chip environment that fluctuating forces due to
magnetic field gradients are smaller (their spectral density scales roughly
with the inverse square of the correlation length). Also when matter-wave
interferometry involves the spatial splitting of a thermal cloud or condensate,
the increase in spatial coherence makes the device more robust against
decoherence from magnetic noise
(see \cite{Cheng1999, Henkel2003, Fermani2006}).
The increase in spatial coherence intimately relates to the reduction 
of heat transfer via fluctuating near fields because the effective number
of channels is inversely proportional to the ``coherence area''. For a more
detailed discussion of this link, see Refs.\cite{Biehs2010,BenAbdallah2010}.

Patch potentials due to adsorbates on the surface are known to add significantly to the electric field noise, relevant for ion traps
\cite{Turchette2000, Dubessy2009} and systems involving precisely tuned 
electric dipole transitions, such as Rydberg states \cite{Carter2011, Tauschinsky2010, Mueller2011}. 
Yet they will not effect the magnetic case. Static patches have no impact on 
the magnetic noise spectrum and magnetic surface-dipoles due to adsorbed 
atoms have only minor impact: 
A static charge trapped at a distance of $1 \unit{\mu m}$ from an atomic-scale electric dipole results in an interaction energy in the order of $20 \unit{MHz}$, while the interaction between two such dipoles gives only  
$1 \unit{kHz}$,
and the magnetic counterpart for two magnetic dipoles of one Bohr magneton
$\mu_B$ at the same distance gives $1 \unit {mHz}$.
More prominent sources of magnetic noise might involve diffusive currents 
confined to a surface layer. Their effect may still be negligible, however, as 
the analysis of Ref.\cite{Henkel2008b} has found.

All of these results imply that nonlocality may be visible at the edge of what is feasible with atom chips. Still, operating a chip trap at short distances
is fundamentally limited by the Casimir-Polder interaction that deforms and breaks the trapping potentials. Alternative setups might, therefore, address the broadening of magnetic transitions spectroscopically, e.g. using evanescent-wave based surface traps as in Ref.\,\cite{Bender2010} or optical tweezers. One may also think of muonic or nuclear magnetic
moments, as used in the experiment of Ref.\,\cite{Suter2004} on spatial dispersion
in superconductors.


\paragraph*{Acknowledgments.} We would like to thank B. Horovitz, F. Intravaia, J. Schiefele, and S. Slama for helpful discussions.
Partial financial support from the German-Israeli Foundation for Scientific Research and Development (GIF), from the European Science Foundation (ESF `Casimir-network'), and from the Deutsche Forschungs\-ge\-mein\-schaft (DFG) is acknowledged.
We thank B. D'Anjou for implementing a numerical model of superconductors during an internship funded by Deutscher Akademischer Austauschdienst (DAAD-RISE).

\begin{appendix}
\section{The magnetic Green's tensor}
\label{app:greens_tensor}

The Green's tensor describes the magnetic field 
\begin{equation}
B_{i}(\vec{r}, \omega ) = \mathcal{H}_{ij}( \vec{r}, \vec{r}', \omega ) 
\mu_j(\omega)	
	\label{eq:def-Hij}
\end{equation}
radiated by a point-like magnetic dipole source placed in $\vec{r}'$.
%
%
Near a single surface, this field consists of a free-space part and a reflected contribution, and therefore 
$\boldsymbol{ \mathcal{H}} = \boldsymbol{ \mathcal{H}}^F +\boldsymbol{ \mathcal{H}}^R$.

%
For coinciding spatial arguments, the imaginary part of the free space 
Green's tensor is given by
%
\begin{eqnarray}
	\label{eq:GF_free_space_1point}
\Im \mathcal{H}^F_{ij}(\vec{r}, \vec{r}, \omega) 
&=& \frac{\mu_0 \omega^3}{6 \pi c^3}\delta_{ij}~.
\end{eqnarray}
A detailed discussion including regularization procedures is given in 
Refs.\,\cite{Vries1998, Cohen-Tannoudji1987}.
The general expression for the reflected Green's tensor reads
\begin{eqnarray}
\boldsymbol{\mathcal{H}}^R(\vec{r},\vec{r'},\omega) &=&
\frac{\mu_0}{4 \pi} \int\limits_0^\infty
\! \frac{p dp}{\kappa} e^{-\kappa |z+z'|}
\left[r_{\rm s}(\omega, p) \boldsymbol{P}+ r_{\rm p}(\omega, p) \boldsymbol{S}\right],\nonumber
\end{eqnarray}
where $\kappa = \sqrt{p^2 - \omega^2/c^2}$, 
$\Re \kappa \ge 0$, $\Im \kappa \le 0$, 
is the propagation constant.
The tensor structure is included in
\begin{eqnarray}
\boldsymbol{P} &=& \left(\begin{array}{ccc}
	\kappa^2\left(J_0(p \rho) - \frac{J_1(p \rho)}{p \rho} \right) & 0 & 
	- p \kappa J_1(p \rho) \\
	0 & \kappa^2 \frac{J_1(p \rho)}{p \rho} & 0 \\
	p \kappa J_1(p \rho) & 0 & p^2 J_0(p \rho)
	\end{array}\right)\nonumber\\
\boldsymbol{S} &=&  \frac{\omega^2}{c^2} \left(\begin{array}{ccc}
	\frac{J_1(p \rho)}{p \rho}& 0 & 0 \\
	0 & J_0(p \rho) - \frac{J_1(p \rho)}{p \rho}  & 0 \\
	0 & 0 & 0	\end{array}\right)~,\nonumber
\end{eqnarray}
where $J_n(x)$ indicate Bessel functions of the first kind. The relative
separation along the surface has length 
$ \rho = [ |\vec{r} -\vec{r}'|^2  - |z-z'|^2]^{1/2}$ and points along the $x$-axis.
In the limit $\vec{r}  \to \vec{r'} $, we obtain the one-point reflected Green's tensor \eqref{eq:mag_greentensor} that depends only on the distance from the surface.
It is easy to check that the free-space contribution~(\ref{eq:GF_free_space_1point})
is typically negligible at distances from a conducting surface smaller
than the vacuum wavelength $c/\omega$,
and we can drop the superscript ${}^R$.

\section{Nonlocal reflectivities}
\label{app:reflectivity}

\subsection{Reflection coefficients}

For $p, s$-polarized incident waves, reflectivities are given by \cite{Ford1984}
\begin{align}
\label{eq:reflectivity_impedance}
r_{\rm p} &= \displaystyle 
\frac{Z_p^0 - Z_p}{Z_p^0 + Z_p},
&r_{\rm s} &= \frac{Z_s - Z_s^0}{Z_s +Z_s^0}
\end{align}
%
where the surface impedances $Z_s, Z_p$ are made dimensionless by
normalizing to the impedance of free space, $\tilde Z = \sqrt{\mu_0 / \varepsilon_0}$.
The well known results from local theory are 
\cite{Jackson1975,Dressel2002} 
\begin{align}
\label{eq:impedance_fresnel}
Z_p^{\rm loc} &= \frac{ \sqrt{\varepsilon(\omega) \omega^2/c^2 -p^2} }{\varepsilon(\omega) \omega / c},
&Z_s^{\rm loc} &= \frac{\omega/c}{ \sqrt{\varepsilon(\omega) \omega^2/c^2 -p^2}}~,
\end{align}
where $\varepsilon( \omega )$ is the dielectric function (we set
$\mu( \omega ) = 1$),
and the reflectivities~(\ref{eq:reflectivity_impedance}) reduce to Fresnel formulas.
The surface impedance on the vacuum side, $Z_p^0, Z_s^0$, is obtained 
by setting $\varepsilon(\omega) = 1$.
Our focus is on metals at low frequencies where in the local limit,
$\varepsilon( \omega ) \omega^2 / c^2 \approx 2 {\rm i} / \delta^2( \omega )$
in terms of the skin depth~(\ref{eq:def_skindepth}), and on
the sub-wavelength limit $p \gg \omega / c$.
This gives an $r_{\rm s}$ coefficient
\begin{equation}
r_{\rm s}^{\rm loc}(\omega, p) 
\approx 
\frac{ {\rm i} p \delta( \omega ) - \sqrt{ 2 {\rm i}  - p^2 \delta^2( \omega ) }
}{ {\rm i} p \delta( \omega ) + \sqrt{ 2 {\rm i}  - p^2 \delta^2( \omega ) } }
~.
%
	\label{eq:local-rs-asymptote}
\end{equation}
The large-momentum asymptote ($p \gg 1/\delta$) is
$r_{\rm s}( \omega, p ) \to {\rm i} / (2 p^2 \delta^2 )$.

\subsection{Impedances for specular and diffuse scattering}

The additional boundary condition that charge carriers undergo specular 
reflection at the inner metal surface can be exploited to extend the
conducting half-space into a fictitious homogeneous medium,
using a similar symmetry for the fields~\cite{Reuter1948, Kliewer1968, Ford1984}. The resulting impedances are 
\begin{eqnarray}
\label{eq:impedance_specular_s}
Z_s^{\rm spec}(\omega, p)&=& \frac{2 i \omega }{c \pi} \int_0^\infty dq \frac{1}{ \varepsilon_t (\omega, \vec{k}) \omega^2/c^2 - k^2} \\
\label{eq:impedance_specular_p}
Z_p^{\rm spec}(\omega, p) &=& \frac{2 i \omega}{c \pi} \int_0^\infty  \frac{dq}{k^2} 
\left[
\frac{p^2}{ \varepsilon_l(\omega, \vec{k}) \omega^2/c^2} \right.+\\
&& + \left.
\frac{q^2}{ \varepsilon_t(\omega, \vec{k}) \omega^2/c^2- k^2}
\right]\nonumber 
~,
\end{eqnarray}
where the medium wave vector is $k^2 = p^2 + q^2$.
The dielectric functions are $\varepsilon_{t,l}
= \varepsilon_b + {\rm i} \sigma_{t,l} / \varepsilon_0 \omega$ with a background polarization $\varepsilon_b$ that drops out in our regime.
If the medium is local, $\varepsilon_{t,l}$ do not depend on $k$, and
a direct calculation of the integrals brings us back to the Fresnel 
impedances \eqref{eq:impedance_fresnel}.

The diffuse reflection of conduction electrons at the surface has been considered 
in Refs. \cite{Reuter1948, Kliewer1970, Foley1975, Halevi1984a}.  
We have used the s-polarized im\-pe\-dance obtained 
in Ref.\,\cite{Foley1975} where the additional boundary condition is
implemented via the so-called dielectric approximation. This means
that in the basic linear current-field relation [Eq.\,(\ref{eq:def-spatial-dispersion})],
the volume integral is restricted to the medium-filled half-space alone.
Solving a Wiener-Hopf equation that follows from the Maxwell equations,
one gets an impedance
\begin{eqnarray}
Z_s^{\rm diff} (\omega, p)&=& \frac{ {\rm i} \pi \omega}{c}
\bigg(\int_0^\infty \!\!\!{\rm d}q 
\log\left[\frac{ q^2 
	}{ k^2 - \varepsilon_t(\omega, \vec{k}) \omega^2/c^2
	}
\right]
\bigg)^{-1}~.
\label{eq:impedance_diffuse_s}
\end{eqnarray}
%
%
This recovers correctly the local limit at all values of $p$. 
However, this is not true for the p-polarized impedance
given in the same work and used in the numerical evaluation in 
Fig.\,\ref{fig:reflectivity_specular_MHz}. 
The violation of 
charge conservation at the surface in the dielectric approximation has 
been discussed in Refs.\,\cite{Flores1977, Flores1979, Foley1975}.
We continue here with the specular boundary condition.

\subsection{Limiting behavior of the reflectivities}
For a metal with specular reflection of charge carriers, Eq.\,\eqref{eq:impedance_specular_s} is rewritten in a dimensionless form by rescaling 
the wavevector $k = \sqrt{p^2 + q^2} = p t$. Making the low-frequency
approximation described before Eq.\,(\ref{eq:local-rs-asymptote}), the
non-local dielectric function becomes [see Eqs.(\ref{eq:def_skindepth}, 
\ref{eq:Boltzmann-Mermin2})]
\begin{equation}
	\varepsilon_t( \omega, \vec{k} ) \approx
	\frac{ 2{\rm i} f_t( 0, {\rm i}/k\ell ) }{ \delta^2( \omega ) }
	\label{eq:asymptote-epsilon-t}
\end{equation}
For simplicity, we drop in the following the frequency arguments and the
redundant argument of the Lindhard function $f_t$. We are left with the
integral
%
\begin{eqnarray}
Z_s^{\rm spec}( p) &=& 
- \frac{2 i \omega }{\pi c } \int_1^\infty dt
\frac{ 1}{\sqrt{t^2-1} \, p t \left[
	1 - 2 {\rm i} f_t( {\rm i} / p \ell t) / (p \delta t)^2 
\right]}
	\nonumber\\
&\approx &  
- \frac{2 i \omega }{\pi c } \int_1^\infty dt
\frac{ 1 + 2 {\rm i} f_t( {\rm i} / p \ell t) / (p \delta t)^2 
	}{\sqrt{t^2-1} \, p t }
\nonumber\\
&=& -\frac{i \omega}{c p}
+ \frac{2 \omega / c}{p^4 \delta^2 \ell}~.
\label{eq:Zs_spec_asymptotic}
\end{eqnarray}
We expanded the denominator in the sub-skin depth regime 
($p \gg 1/\delta, 1/\ell$)
and replaced, in the last step,  the transverse Lindhard function 
[Eq.\,\eqref{eq:Boltzmann-Mermin4}] by its asymptote 
for large $p \ell$:
\begin{equation}
	f_t( {\rm i} / p \ell t ) =
\frac{ 3 \pi  }{ 4 p \ell t } + {\cal O}[(p \ell t)^{-2}]
~.
	\label{eq:large-k-Lindhard-ft}
\end{equation}
This result complies with a similar calculation carried 
out in Ref.\,\cite{Esquivel2004}.

The first term in Eq.\eqref{eq:Zs_spec_asymptotic} corresponds to the
free-space surface impedance in the sub-wavelength limit. The
reflection amplitude\,\eqref{eq:reflectivity_impedance} therefore becomes
\begin{eqnarray}
r_{\rm s}^{\rm spec}( p ) &\approx& 
\frac{Z_s^{\rm spec}( p ) + {\rm i} \omega / ( c p )
	}{
	Z_s^{\rm spec}( p ) - {\rm i} \omega / ( c p )
	} 
	\approx 
\frac{ {\rm i} }{ p^3 \delta^2 \ell}
%
			\label{eq:rs_specular_BM_large_k_app}
\end{eqnarray} 
which gives Eq.\,(\ref{eq:rs_specular_BM_large_k}).
%
 
In the scenario where charge carriers are reflected diffusely rather than specularly
we start from Eq.\,\eqref{eq:impedance_diffuse_s}. 
The same substitution of the integration variable gives
\begin{equation}
Z_s^{\rm diff}( p ) = 
\frac{\displaystyle {\rm i} \pi \omega / ( c p )
	}{ 
\displaystyle
\int_1^\infty \!\!\frac{dt\,t}{\sqrt{t^2-1}}
\log\left[\frac{t^2-1}{t^2 - 
2 {\rm i} f_t( {\rm i} / p \ell t ) / (p \delta t)^2 }
\right]}~.
\end{equation}
The integrand is expanded  in  $1/(p \delta)^2$, and the integration can be performed explicitly
\begin{eqnarray}
&&\int_1^\infty \frac{dt\,t}{\sqrt{t^2-1}}
\log\left[\frac{t^2-1
	}{ t^2 - 
	2 {\rm i} (p \delta t)^{-2} f_t( {\rm i} / p \ell t ) 
}\right] 
\nonumber\\
&\approx&
\int_1^\infty \frac{dt\,t}{\sqrt{t^2-1}}
\left\{
\log\left[\frac{t^2-1}{t^2}\right] 
+\frac{2 {\rm i} f_t( {\rm i} / p \ell t )  }{(p \delta t)^2}
\right\}
\nonumber\\
&= & - \pi + \frac{ 3\pi {\rm i} }{ 2 p^3 \delta^2 \ell }
\qquad
\mbox{ if } p \ell \gg 1
~.
\end{eqnarray}
%
The resulting impedance reads  
\begin{eqnarray}
Z_s^{\rm diff}( p ) 
&\approx& 
- \frac{i \omega }{c p} + \frac{3 \omega / c }{ 2 p^4 \delta^2 \ell }
~.
\end{eqnarray}
This differs from Eq.\,\eqref{eq:Zs_spec_asymptotic} only by a factor $3/4$ 
in the second term, 
%
so that the reflection amplitude is smaller by this number compared to the
specular boundary condition, Eq.\,(\ref{eq:rs_specular_BM_large_k_app}).


\end{appendix}

\end{document}